\documentclass{article}
\usepackage{epsfig}
\title{Photoluminescence and photoluminescence excitation studies of lateral 
size 
effects in Zn$_{1-x}$Mn$_x$Se/ZnSe quantum disc samples of different radii}

\author{%
P. J. Klar, D. Wolverson and J. J. Davies\\
\textit{School of Physics, University of East Anglia}\\
\textit{Norwich NR4 7TJ, United Kingdom}\\
W. Heimbrodt and M. Happ\\
\textit{Institut f\"ur Physik, Humboldt-Universit\"at zu Berlin}\\
\textit{Invalidenstra\ss{}e 110, D-10115 Berlin, Germany}\\
T. Henning\\
\textit{Applied Solid State Physics, University of G\"oteborg and}\\
\textit{Chalmers University of Technology, S-41296 G\"oteborg, Sweden}
}

\date{cond-mat/9803208}

\begin{document}
\maketitle
\begin{abstract}
Quantum disc structures (with diameters of 200\,nm and 100\,nm) were 
prepared from a 
Zn$_{0.72}$Mn$_{0.28}$Se/ZnSe single quantum well structure by electron beam 
lithography followed by 
an etching procedure which combined dry and wet etching techniques. The 
quantum disc 
structures and the parent structure were studied by photoluminescence 
and photoluminescence 
excitation spectroscopy. For the light-hole excitons in the quantum 
well region, shifts of the 
energy positions are observed following fabrication of the discs, 
confirming that strain 
relaxation occurs in the pillars. The light-hole exciton lines also 
sharpen following disc 
fabrication: this is due to an interplay between strain effects 
(related to dislocations) and the 
lateral size of the discs. A further consequence of the small lateral 
sizes of the discs is that the 
intensity of the donor-bound exciton emission from the disc is found to 
decrease with the disc 
radius. These size-related effects occur before the disc radius is 
reduced to dimensions 
necessary for lateral quantum confinement to occur but will remain 
important when the discs 
are made small enough to be considered as quantum dots.
\end{abstract}

%I.     INTRODUCTION
\section{Introduction}
        The development of etching methods for II-VI semiconductor 
heterostructures which 
allow a further reduction of dimensionality towards quantum dots 
(Q-dots) and quantum wires 
(Q-wires) is currently of interest. So far, three different etching 
techniques have been used: ion 
beam etching (IBE) \cite{lit01,lit02}, reactive ion etching (RIE)
\cite{lit03} and wet-chemical 
etching (WCE) \cite{lit04,lit05}. The 
anisotropy of dry etching processes such as RIE or IBE allows the 
fabrication of patterns with 
a small spacing and of considerable etch depth. However, nanostructures 
fabricated using IBE 
show considerable surface damage (up to depths of about 30\,nm)
\cite{lit06}. In 
contrast, the surface 
damage induced by WCE is small: recently, the preparation by this 
method of high-quality 
Zn$_{1-x}$Cd$_x$Se/ZnSe Q-dot and Q-wire structures down to lateral sizes of 
20\,nm with a high 
photoluminescence (PL) quantum efficiency was reported and, in that 
work, the observation of 
quantum confinement effects via PL was possible \cite{lit05}. The major 
disadvantage of WCE is that, 
because of the mainly isotropic etch process, only low-density patterns 
of Q-dots and Q-wires 
can be obtained. From the point of view of future applications, it is 
desirable to fabricate large 
areas of high-quality nanostructures with high-density patterns by 
methods which combine the 
advantages of dry and wet-chemical etching. Attempts in this direction 
have been made by 
Gurevich et al., who combined RIE and WCE
\cite{lit06} and by Gourgon et al., who 
used anodic 
oxidation after IBE to remove the damaged surface layer \cite{lit07}.
\par
We report here on an etching technique which combines Ar$^{2+}$ IBE 
and WCE. Using this 
novel combination of techniques, we prepared large areas 
($3.2\,\mathrm{mm}
\times 3.2\,\mathrm{mm}$) of 
Zn$_{0.72}$Mn$_{0.28}$Se/ZnSe pillars containing single quantum wells of 
uniform diameters of either 
200\,nm or 100\,nm, with area densities respectively of 1:4 and 1:16 in 
the pattern. The lateral 
dimensions of the quantum layers are too large for lateral quantum 
confinement effects to be 
important and we therefore refer to the structures as quantum discs 
(rather than quantum 
dots). In studying these discs, we find that there are significant size 
effects of a \emph{non-quantum} 
nature that must be taken into account and which must be fully 
elucidated if the behaviour of 
quantum dots themselves is also to be understood.
\par
Nanostructures from the Zn$_{1-x}$Mn$_x$Se/ZnSe system are of 
interest not only because of 
their close relation to the wide-bandgap II-VI materials used in 
blue-green optoelectronic 
devices, but also because of the variety of additional effects that 
arise from the unique 
magnetic properties of dilute magnetic semiconductors (DMS). Although 
the effects of 
magnetic field are not explicitly part of the present work, we shall 
make use of spectral 
changes induced by magnetic fields in order to identify certain 
transitions. The usefulness of 
magneto-optical experiments on DMS quantum structures to address 
problems of 
nanofabrication as well as fundamental physical problems has been 
demonstrated elsewhere \cite{lit08}.

%II. SAMPLE PREPARATION
\section{Sample preparation}
The Zn$_{0.72}$Mn$_{0.28}$Se/ZnSe single quantum well (SQW) structure 
used for the 
nanofabrication of the Q-dots was grown on an almost exactly (100) 
oriented GaAs substrate 
in a DCA 350 MBE system equipped with effusion cells for Zn, Cd, Mn and 
Se. The growth 
was carried out in a phase-locked epitaxial mode based on RHEED
\cite{lit09}. The 
SQW structure was 
deposited on a ZnSe buffer layer of 730\,nm thickness and consisted of a 
505\,nm thick barrier 
layer followed by the quantum well layer of 5.1\,nm and a 20\,nm barrier 
as a capping layer 
(figure~1).
\begin{figure}
\centering
\epsfig{file=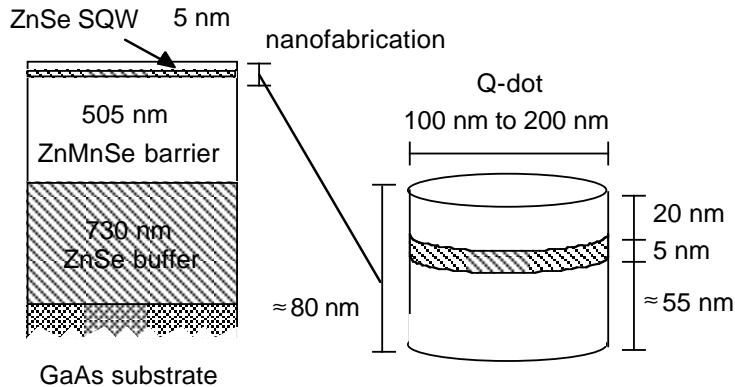,width=0.8\textwidth}
\caption{Schematic diagrams of the 
Zn$_{0.72}$Mn$_{0.28}$Se/ZnSe single 
quantum well structure (left) 
and of a pillar with quantum disc (right). The pillars are etched only 
to a depth of about 80\,nm 
from the surface into the sample in the nanofabrication process. The 
layer thicknesses of the 
parent structure and the dimensions of the pillars are indicated.}
\end{figure} 
\par
        The nanofabrication process of the Q-dot samples was based on 
the eight step 
technique described in detail earlier
\cite{lit02}. Only small modifications were 
made for the preparation 
of the 200\,nm Q-dot sample. The dose in the electron beam lithography 
process was reduced 
to 205\,$\mu$C/cm$^2$ and etching was carried out by Ar$^{2+}$ IBE through the 
quantum well into the 
lower lying barrier to a depth of about 80\,nm. The first fabrication 
steps for the 100\,nm Q-dot 
sample were the same as for the 200\,nm Q-dot sample. However, to reduce 
the diameter of the 
Q-dots to 100\,nm, wet-chemical etching was used as an additional step 
before the removal of 
the Ti-protection mask. The etchant used was a solution of bromine and 
ethylene glycol with a 
solution ratio of Br$_2$:HOCH$_2$CH$_2$OH of 2:1000
\cite{lit04}. Figure~2 shows an SEM 
image of the 
100\,nm Q-dot sample before the removal of the Ti-protection mask.
\begin{figure}
\centering
\epsfig{file=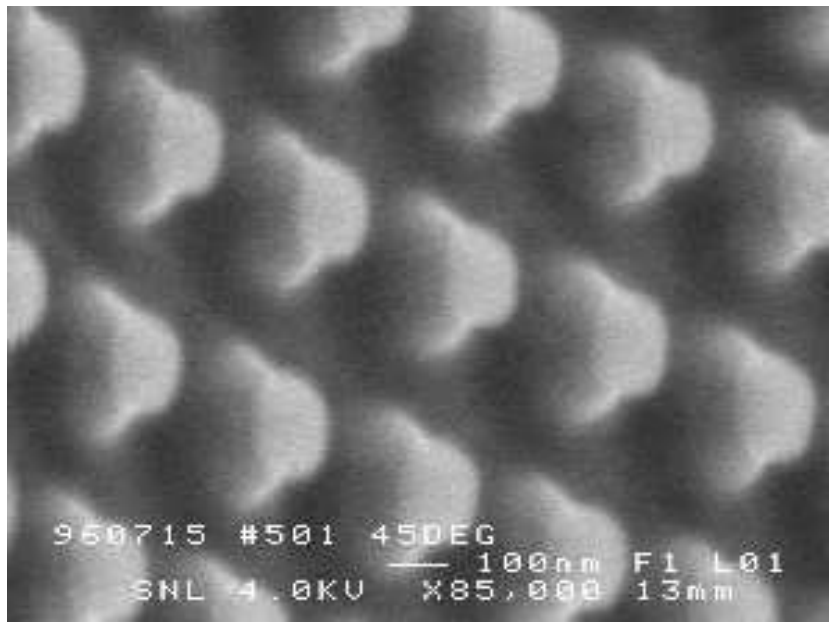,width=0.9\textwidth}
\caption{Image of the 100\,nm Q-disc pillars taken by scanning electron 
microscopy after 
wet-chemical etching and before removal of the titanium protection 
mask. The scale of the 
figure is indicated by the horizontal bar.}
\end{figure}

%III. RESULTS AND DISCUSSION
\section{Results and Discussion}
The PL and the photoluminescence excitation (PLE) experiments 
were carried out with 
the sample in liquid helium at 2 \,K. A tunable dye laser (Stilbene\,3) 
pumped by the UV emission 
of an Ar$^{2+}$ laser was used as excitation source and a single grating 
spectrometer equipped with 
a charge-coupled device detector was used for detection. The PL spectra 
were recorded 
directly, whereas the PLE spectra were derived from sets of complete PL 
spectra taken at 
different finely-spaced excitation energies. The magneto-optical 
measurements were carried 
out by means of a superconducting magnet system. Spectra were taken in 
the Faraday 
configuration (with the layer plane of the specimens normal to the 
magnetic field) for $\sigma^+$ and $\sigma^-$
circularly polarised light at magnetic fields up to 7~Tesla.
\par
\begin{figure}
\centering
\epsfig{file=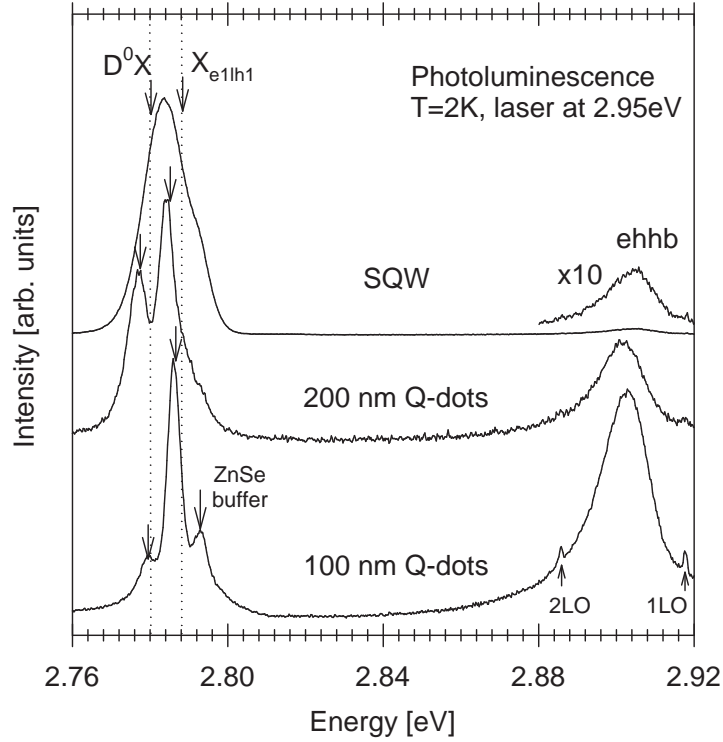,width=0.8\textwidth}
\caption{Photoluminescence spectra of the parent 
Zn$_{0.72}$Mn$_{0.28}$Se/ZnSe 
structure and the 
200\,nm and 100\,nm Q-disc samples. The spectra are shifted vertically 
for clarity. D$^0$X and 
X$_\mathrm{e1lh1}$ denote the donor-bound and the free exciton emission bands 
respectively from the 
quantum well layer in each specimen and, in the lowest spectum, two 
bands arising from 
Raman scattering of phonons are also indicated (1LO and 2LO).}
\end{figure}
        Figure~3 depicts a comparison of the PL spectra of the three 
specimens. All the spectra 
were obtained with the energy of the excitation laser light above the 
bandgap of the 
Zn$_{0.72}$Mn$_{0.28}$Se barrier. 
At the right-hand end of each spectrum a signal 
appears at about 2.905\,eV: 
the shift in the position of this signal when a magnetic field is 
applied enables it to be 
identified as being due to a heavy-hole barrier (ehhb) exciton (the 
large shift in energy of such 
transitions is caused by the so-called giant enhancement of the applied 
magnetic field by the 
exchange interaction between the charge carriers and the magnetic ions 
in the dilute magnetic 
material). In the left-hand part of the PL spectrum from the quantum 
well layer (upper 
spectrum) there appears a signal made up of two overlapping bands. 
Again, when a magnetic 
field is applied, both these bands shift in energy and can be 
attributed to recombination 
transitions in the quantum well involving the free light-hole exciton, 
denoted by X$_\mathrm{e1lh1}$, and, at 
slightly lower energy, a donor-bound exciton, denoted as D$^0$X (the field 
induced shifts in these 
transitions are due to penetration of the well wavefunctions into the 
barrier regions, which are 
magnetic). 
\par
        When the structures are etched to form the quantum discs, there 
are no significant 
changes in the emission from the Zn$_{0.72}$Mn$_{0.28}$Se barriers (the 
successive increases in intensity 
are due simply to the increase, through uncovering, of the area of the 
barrier directly exposed 
to the excitation). In contrast, there are several interesting changes 
in the PL from the ZnSe 
well.  For the 200\,nm discs, the  D$^0$X and X$_\mathrm{e1lh1}$ lines are both shifted 
towards lower energy by 
about 4\,meV relative to the corresponding lines in the unetched 
material and are sharpened. 
When the disc radius is reduced to 100\,nm, the exciton lines remain 
sharp and shift partly (by 
about 2\,meV), but not completely, back towards their positions in the 
original layer. A further 
noticeable feature is the marked decrease in the intensity of the D$^0$X 
emission relative to that of 
the X$_\mathrm{e1lh1}$ in going from the original layer to the 200\,nm Q-discs and 
then to those of 100\,nm 
diameter, for which the D$^0$X emission has almost disappeared. The 
emission band at 2.792\,eV 
in the PL spectrum of the 100\,nm Q-dot sample originates from the ZnSe 
buffer, as is readily 
confirmed by its lack of sensitivity to an applied magnetic field.
\par
        Further information about the excitonic states of the samples 
can be obtained from the 
PLE spectra (which were taken by monitoring the low energy side of the 
quantum well PL). 
Before we compare the three samples, it is useful to look at the 
evolution of the PLE spectra 
with magnetic field, which helps to identify the peaks unambiguously. 
\begin{figure}
\centering
\epsfig{file=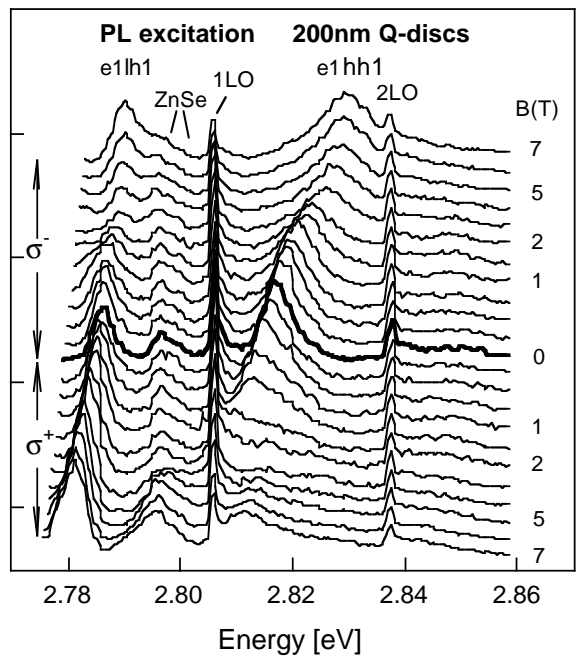,width=0.8\textwidth}
\caption{Photoluminescence excitation spectra of the 200\,nm Q-discs 
sample at magnetic 
fields from 0 to 7 Tesla in the Faraday configuration with $\sigma^-$ and 
$\sigma^+$ 
circularly polarised light. 
The magnetic field step size was 0.25\,T from 0 to 1\,T, and 1\,T beyond 
that; the zero-field 
spectrum is indicated by the heavier line.}
\end{figure}
In figure~4, the excitation 
spectra of the 200\,nm Q-discs are depicted as an example. The sharp 
lines are due to one- and 
two-phonon LO phonon Raman scattering and do not move in an external 
magnetic field. 
There are also no shifts of the bands lying somewhat below 2.8\,eV and 
these are attributed, 
therefore, to the ZnSe buffer layer. However, a clearly observable 
shift is seen for both the 
remaining bands, which are therefore ascribed to the e1lh1 and e1hh1 
well states. The shift 
with magnetic field is caused by the exchange interaction between the 
well excitons and the 
localised Mn spins in the barrier. Shifts of about 4\,meV and 13\,meV to 
higher energies in case 
of $\sigma^-$ circularly polarised light are typically observed for the e1lh1 
and e1hh1 excitons 
respectively in ZnSe wells of width about 5\,nm between 
Zn$_{0.72}$Mn$_{0.28}$Se 
barriers. The behaviour 
of the PLE bands in the case of $\sigma^+$ circularly polarised light is more 
complicated and a detailed 
discussion is beyond the scope of this paper: briefly, a transition of 
the band alignment from 
type-I to type-II leads to the disappearance of the e1hh1 PLE peak at a 
magnetic field of about 
1.5~Tesla. 
\par
\begin{figure}
\centering
\epsfig{file=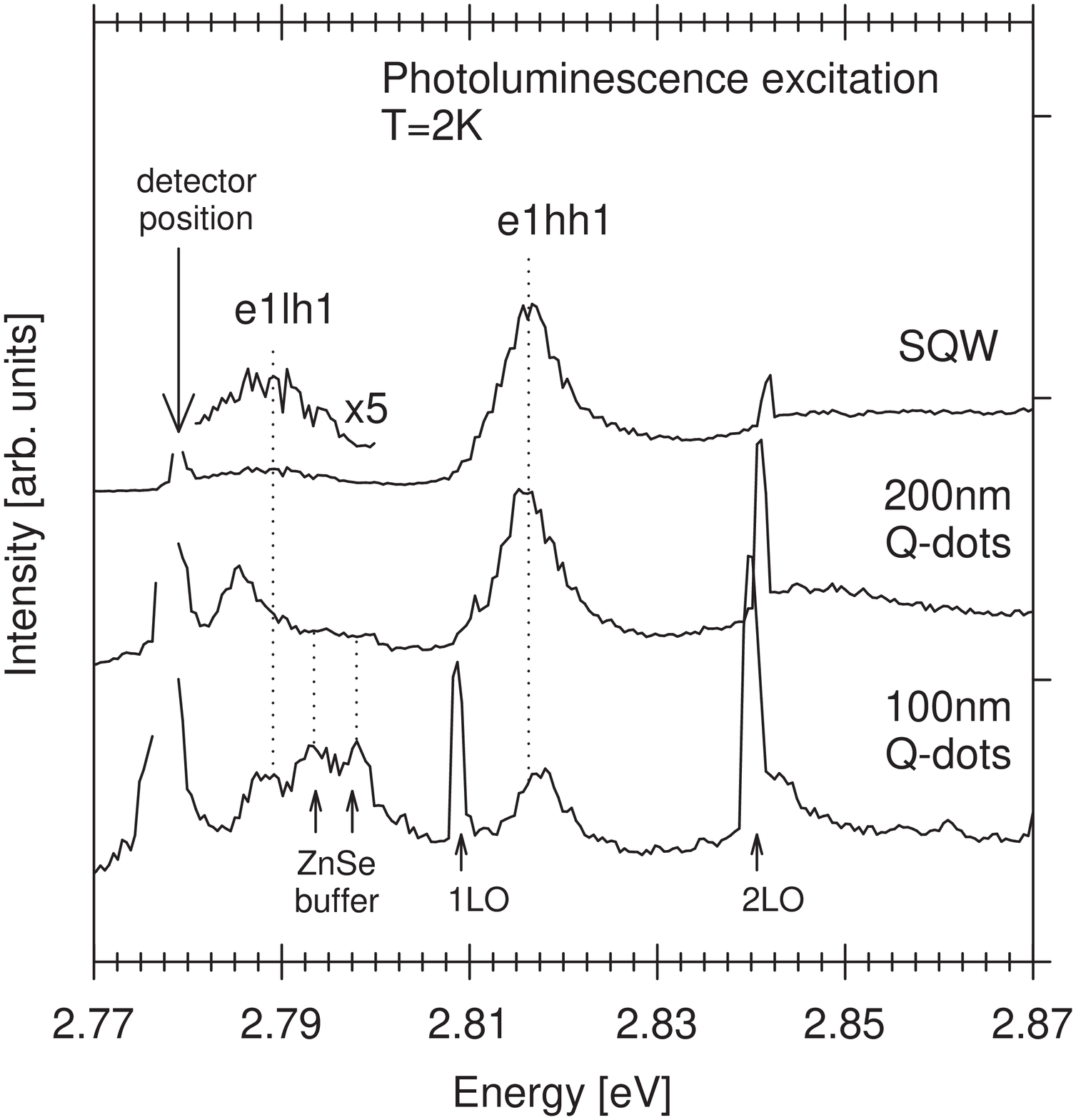,width=0.8\textwidth}
\caption{Photoluminescence excitation spectra of the parent 
Zn$_{0.72}$Mn$_{0.28}$Se/ZnSe structure 
and the 200\,nm and 100\,nm Q-disc samples. The spectra are shifted 
vertically for clarity. The 
quantum well excitonic heavy-hole and light-hole transitions are 
labelled e1hh1 and e1lh1. 
Two lines arising from LO phonon Raman scattering are indicated by 1LO 
and 2LO 
respectively.}
\end{figure}
        Let us now compare the excitation spectra (figure~5) of the 
three specimens at zero 
field. For the original, unetched layer, two bands are observed that 
can be attributed to 
excitonic transitions. The first of these bands, at about 2.815\,eV, is 
assigned to the lowest 
heavy-hole quantum well exciton (e1hh1), this interpretation being 
consistent with the strain in 
the layer (see below). The second band is much broader and much weaker 
and, because it rests 
on a broad PLE background, the position of its peak intensity is 
difficult to determine; 
nevertheless, it can be assigned to the e1lh1 exciton of the well. In 
the spectra of the Q-discs, 
the e1lh1 excitonic bands are sharper and their positions are close to 
those of the 
corresponding emission bands; as in the PL spectra, their peak 
positions are shifted first 
towards lower energy (for the 200\,nm Q-discs) and then back by about 
2\,meV (for the 100\,nm 
Q-discs).  In contrast, the position of the e1hh1 band remains the same 
and its width is 
unaltered; this observation is readily accounted for since the heavy 
hole exciton transitions are 
expected to be much less sensitive to strain than are those of the 
light hole excitons. 
\par
        For the original SQW layer, when the detection energy for the 
PLE is changed, the 
bands attributed to the heavy-hole excitons (e1hh1) remain fixed in 
energy (as expected). The 
behaviour of the band attributed to the light-hole exciton (e1lh1) is, 
however, rather 
complicated. Firstly, we remark that the intensity of this band 
relative to that of the heavy-hole 
band is smaller than expected (the oscillator strengths for the 
corresponding absorption bands 
are in the ratio 1:3 and this ratio is approached in the case of the 
quantum discs). Secondly, the 
peak position appears not to be independent of the detection 
wavelength. These observations 
suggest that the weak signal shown at 2.789\,eV in PLE spectrum for the 
SQW (figure~5) is 
only part of a broader band. There therefore appears to be a broadening 
mechanism that affects 
the light-hole transitions in the SQW which is absent in the quantum 
discs. As discussed later, 
we believe this mechanism to be strain-related, so that the heavy-hole 
transitions would be 
affected less severely. In the particular case of the 100\,nm Q-discs, 
two additional bands are 
observed, at 2.793\,eV and 2.798\,eV, both originating from the ZnSe 
buffer layer. The 
assignments of the signals in the PLE spectra are confirmed by their 
behaviour in a magnetic 
field (these measurements will be discussed elsewhere).
\par
        Before the discussion of the observations made in figures~3 and 
5, it is helpful to recall 
the structure (figure~1) of the initial Zn$_{0.72}$Mn$_{0.28}$Se/ZnSe SQW 
sample and that of the Q-disc 
samples after the nanofabrication. The strain in the SQW is determined 
by the thicknesses of 
the different layers that make up the structure. The thicknesses of the 
ZnSe buffer layer and the 
lower Zn(Mn)Se barrier ensure that the region of the buffer, barrier 
and SQW is relaxed 
relative to the GaAs substrate, the SQW region itself being strained to 
a lattice constant 
intermediate between that of ZnSe and Zn(Mn)Se. Since the Zn(Mn)Se 
lattice constant is 
slightly greater than that of ZnSe, the SQW is therefore subjected to 
biaxial tensile strain, to 
which a further contribution is made as a result of the differential 
thermal contraction that 
occurs between the GaAs substrate and the ZnSe and Zn(Mn)Se layers when 
the specimen is 
cooled. These strain effects, together with the effects of the quantum 
confinement in the well 
of thickness 5.1\,nm, determine the energies of the excitonic 
transitions in the well region of the 
unetched structure.
\par
        When the quantum well structure is etched to form the pillars, 
the strain will change as 
elastic relaxation occurs. Since the Zn$_{0.72}$Mn$_{0.28}$Se has a greater 
lattice constant than the ZnSe 
and since the quantity of Zn$_{0.72}$Mn$_{0.28}$Se in the pillar 
greatly exceeds 
that of the ZnSe, the 
tensile strain in the quantum well is expected to increase when the 
discs are formed. Such an 
increase in tensile strain  displaces the D$^0$X and X$_\mathrm{e1lh1}$ lines to lower 
energies, as shown in the 
central spectra of figures~3 and 5.
\par
This shift of the D$^0$X and X$_\mathrm{e1lh1}$ lines would be expected at 
first sight to become even 
larger as the disc radius is further decreased. Experimentally, this is 
clearly not the case and 
this suggests that when the disc diameter becomes too small, a further 
elastic relaxation takes 
place in which the quantum well is no longer fully constrained to the 
lattice constant of the 
(much thicker) barrier. For thin layers in free-standing Q-wire 
structures, such relaxation 
effects have recently been calculated and compared with experimental 
observations by Niquet 
et al. \cite{lit10}, who find that for aspect ratios of wire 
width to well width 
of less than ten, significant 
relaxation of the strain occurs in the regions of the well. In the case 
of quantum wells in pillar 
structures of the form studied in the present paper, relaxation of the 
strain in the well would be 
expected to occur at larger aspect ratios than for wires. In the 100\,nm 
discs, the aspect ratio is 
of the order of 20 (see figure~1) and the experimentally observed 
relation is therefore entirely 
reasonable. 
\par
        With decreasing disc size we observe a reduction of the PL 
linewidths. In the quantum 
well PL emissions of the 200\,nm Q-discs and of the 100\,nm Q-discs the 
D$^0$X and X$_\mathrm{e1lh1}$ bands 
are resolved, whereas for the SQW sample they overlap. There are three 
explanations for this 
effect that can be considered. The first is that the broadening in the 
original layer is due to 
fluctuations in the alloy composition of the barrier material (a 
similar broadening has been 
observed recently in single In$_{1-x}$Ga$_x$As/GaAs quantum discs and has 
been explained by 
fluctuations in alloy composition of the well
\cite{lit11}): however, this 
mechanism can be excluded in 
the present case, since, for a laser spot of about 1\,mm$^2$, we probe 
simultaneously more than six 
million discs, in which a distribution of alloy concentrations would 
still be present. A second 
explanation is that the linewidth in the original SQW is due to 
fluctuations in the well width: 
again, this mechanism can be excluded since such fluctuations would 
remain after the 
formation of the discs.
\par
        The third explanation (and the one that we favour) is that the 
linewidth in the original 
SQW is determined mainly by the strain fields associated with 
dislocations. To quantify this 
explanation, we shall assume a typical value for the dislocation 
density of 10$^7$\,cm$^{-2}$, as observed 
in similar samples
\cite{lit12} (implying a mean separation of $3 \cdot10^4$\,\AA{}). At a 
distance $r$ from a 
dislocation, the strain due to it may be estimated by  
$b/(2\pi r)$, where $b$ is the 
Burgers vector for 
the dislocation. The implications of this for the PL linewidth before 
nanofabrication can be seen 
by noting that the X$_\mathrm{e1lh1}$ exciton in 
ZnSe shifts in energy by 4\,meV for 
a change in strain of 
0.5\,\% \cite{lit13}. 
If, for the purposes of a simple estimate, we consider regions 
of the layer as 
unaffected by a dislocation when the local X$_\mathrm{e1lh1}$ 
energy is shifted by 
less than 0.1\,meV then a 
value of $r$ may be found which defines the boundary separating the 
regions affected and 
unaffected by that dislocation (around 7000\,\AA{}). Comparing the area of 
affected regions with 
the mean area per dislocation (of $9 \cdot 10^8$\,\AA{}$^2$) 
shows that in about 20\,\% 
of the epilayer, the light 
hole exciton energy is shifted by at least 0.1\,meV due to the presence 
of the dislocation-related 
strain fields. The broadening of the X$_\mathrm{e1lh1}$ emission is therefore 
expected to be significant 
before nanofabrication. When discs are formed by etching, the 
probability of a given disc 
containing a dislocation is, even for a diameter of 200\,nm, very small 
(for a dislocation density 
of 10$^7$\,cm$^{-2}$ the probability is less than 1\,\%). 
Under these assumptions, 
the quantum discs would 
therefore be expected to be uniformly in the same strain state, giving 
rise to the observed 
narrowing of the light-hole excitonic bands in the optical spectra (the 
heavy-hole lines are 
already narrow since they are less sensitive to strain and are not 
affected significantly in this 
way).
\par
        The observation that the intensity of the PL emission of the 
bound exciton D$^0$X 
decreases more strongly than the free exciton emission X$_\mathrm{e1lh1}$ with 
decreasing Q-disc size is, at 
first sight, unexpected, but can be accounted for by a similar 
size-related argument, as depicted 
schematically in figure~6. 
\begin{figure}
\centering
\epsfig{file=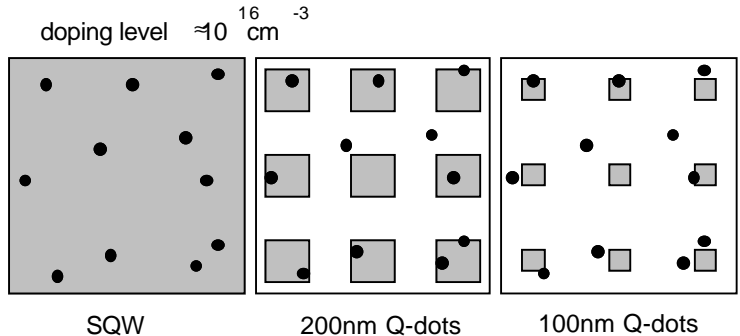,width=0.8\textwidth}
\caption{Illustration of the increase of the number of quantum discs 
without donor ions in the 
quantum well layer with decreasing disc radius. The black circles 
represent the donor ions and 
the grey areas represent schematically the remaining quantum well layer 
after etching in the 
three specimens. Note that the diagram is drawn approximately to scale.}
\end{figure}
The SQW sample is unintentionally doped 
n-type with an upper limit 
of the doping level being 10$^{16}$\,cm$^{-3}$. 
If the ZnSe quantum well region of 
a Q-disc of diameter $d$ 
is approximated by a box of $5\,\mathrm{nm} \times
 d \times d$, a doping level of 10$^{16}$\,cm$^{-3}$ 
results in an average of 2 
donor ions per ZnSe well in a 200\,nm Q-disc and of 0.5 donor ions per 
ZnSe well in a 100\,nm 
Q-disc. Statistically, the number of Q-discs without donor ions in the 
ZnSe well increases with 
decreasing Q-disc size. For this argument to hold, it is essential 
that, in the ZnSe quantum well 
of the unetched sample, the cross-sectional areas of the donor ions for 
exciton capture overlap 
or, at least, that the cross-sectional area for exciton capture of an 
individual donor ion is bigger 
than the lateral dimension of the Q-discs in the etched specimen i\,e. 
bigger than 100\,nm to 
200\,nm. Most of the photo-created excitons in the unetched specimen 
will then reach a donor 
ion before they decay radiatively. The mean distance between donors is 
about 450\,nm for a 
density of 10$^{16}$\,cm$^{-3}$. 
Most of the excitons which contribute to the 
quantum well PL emission 
are created in the Zn$_{0.72}$Mn$_{0.28}$Se barriers and migrate to the ZnSe 
well. These excitons 
therefore have a non-zero momentum and can travel long distances in the 
quantum well before 
they decay radiatively. Thus, the decrease of the bound exciton 
emission D$^0$X relative to the 
free exciton emission X$_\mathrm{e1lh1}$ can be consistently explained by an 
increase of the number of 
Q-discs without donor ions in the ZnSe well region with decreasing 
Q-disc size and can be 
considered as a genuine size effect.

%IV. CONCLUSIONS
\section{Conclusions}
        The combination of Ar$^{2+}$ ion beam etching and wet-chemical 
etching offers a promising 
way for the fabrication of good quality quantum disc structures with 
high pattern densities. The 
optical studies of the two quantum disc samples (100\,nm and 200\,nm in 
diameter) prepared by 
this method and of their 
Zn$_{0.72}$Mn$_{0.28}$Se/ZnSe SQW parent structure 
reveal a line-narrowing 
of the light-hole excitonic bands in both the PL and PLE spectra, 
together with an 
enhancement of the intensity of the free exciton relative to the 
donor-bound exciton in the 
quantum discs. Both these observations are a consequence of the small 
lateral size of the discs 
but are not related to quantum confinement in the plane of the disc. 
The shifts in the positions 
of the light-hole exciton lines are related to elastic relaxation that 
occurs during the formation 
of the pillars and suggest that pillars 100\,nm in height and with a 
diameter of 200\,nm are 
already small enough to behave as free-standing structures but that 
when the ratio of disc 
radius to pillar height becomes too small, further relaxation occurs, 
as pointed out for quantum 
wire structures
\cite{lit10}. Because of the size effect that results in the 
probability of a dislocation being 
present in a given pillar being very small, all quantum discs are 
relaxed to a similar strain state. 
An important aspect of the work is the insight into effects of 
nanofabrication that can be gained 
by studying structures in an intermediate size regime before the 
additional effects caused by 
lateral quantum confinement complicate the behaviour. The knowledge 
gained by the study of 
quantum discs is therefore an important step towards the understanding 
of structures in which 
additional lateral confinement effects do indeed arise.

%ACKNOWLEDGEMENTS
\section*{Acknowledgements}
        We gratefully acknowledge the financial support of the British 
Council, Deutscher 
Akademischer Austauschdienst and of the Engineering and Physical 
Sciences Research Council 
(GR/K04859). PJK thanks the University of East Anglia for a research 
studentship. We also 
thank N. Hoffmann and J. Griesche for the growth of the SQW sample from 
which the discs 
were fabricated.


\begin{thebibliography}{99}
\bibitem{lit01} L. S. Dang, C. Gourgon, N. Magnea, H. Mariette and 
C. Vieu, Semicond. 
Sci. Technol. \textbf{9}, 
1953 (1994)
\bibitem{lit02} P. J. Klar, D. Wolverson, D. E. Ashenford, 
B. Lunn and T. Henning, 
Semicond. Sci. Technol. 
\textbf{11}, 1863 (1996)
\bibitem{lit03} M. A. Foad, C. D. W. Wilkinson, C. Dunscomb and 
R. H. Williams, Appl. 
Phys. Lett. \textbf{60}, 
2531 (1992)
\bibitem{lit04} G. Bacher, M. Illing, A. Forchel, D. Hommel, 
B. Jobst and G. Landwehr, 
phys. stat. sol.(b) 
\textbf{187}, 371 (1995)
\bibitem{lit05} M. Illing, G. Bacher, T. K\"ummell, A. Forchel, 
T.~G. Andersson, D. 
Hommel, B. Jobst and G. 
Landwehr, Appl. Phys. Lett. \textbf{67}, 124 (1995)
\bibitem{lit06} S. A. Gurevich, O. A. Lavrona, N. V. Lomasov, 
S. I. Netserov, V. I. 
Skopina, 
E. M. Tanklevskaya, V. V. Travnikov, A. Osinsky, Y. Qiu, H. Temkin, M. 
Rabe and 
F. Henneberger, to be published in Electronic Letters (1997)
\bibitem{lit07} C. Gourgon, L. S. Dang, H. Mariette, C. Vieu and 
F. Muller, Appl. 
Phys. Lett. \textbf{66}, 1635 
(1995)
\bibitem{lit08} P. J. Klar, D. Wolverson, J. J. Davies, B. Lunn, 
D. E. Ashenford and 
T. Henning , 23rd Int. 
Conf. on the Physics of Semiconductors, eds. M Scheffler and R 
Zimmermann, vol. 2 
1485 (World Scientific, 1996)
\bibitem{lit09} J. Griesche, N. Hoffmann and K. Jacobs, 
J. Cryst. Growth \textbf{138}, 59 (1994)
\bibitem{lit10} Y. M. Niquet, C. Priester and H. Mariette, 
Phys. Rev. B \textbf{55}, R7387 
(1997)
\bibitem{lit11} R. Steffen, A. Forchel, T. L. Reinecke, 
T. Koch, M. Albrecht, J. 
Oshinowo and F. Faller, 
Phys. Rev. B \textbf{54}, 1510 (1996)
\bibitem{lit12} Z. H. Yu, S. L. Buczkowski, N. C. Giles and 
T. H. Myers, Appl. Phys. 
Letts. \textbf{69}, 82 (1996)
\bibitem{lit13} H. Mayer, U. R\"ossler, K. Wolf, A. Elstner, 
H. Stanzl, T. Reisinger 
and W. Gebhardt, Phys. 
Rev. B \textbf{52}, 4956 (1995)
\end{thebibliography}
\end{document}